\documentclass[aps,pra,a4paper,twocolumn]{revtex4}
\usepackage{graphicx,amsmath}

\begin{document}

\title{What are single photons good for?}
   
\author{Nicolas Sangouard}
\author{Hugo Zbinden}
\affiliation{Group of Applied Physics, University of Geneva, 1211 Geneva 4, Switzerland}

\begin{abstract}
In a long-held preconception, photons play a central role in present-day quantum technologies. But what are sources producing photons one by one good for precisely? Well, in opposition to what many suggest, we show that single-photon sources are not helpful for point to point quantum key distribution because faint laser pulses do the job comfortably. However, there is no doubt about the usefulness of sources producing single photons for future quantum technologies. In particular, we show how single-photon sources could become the seed of a revolution in the framework of quantum communication, making the security of quantum key distribution device independent or extending quantum communication over many hundreds of kilometers. Hopefully, these promising applications will provide a guideline for researchers to develop more and more efficient sources, producing narrowband, pure and indistinguishable photons at appropriate wavelengths.
\end{abstract}
\date{\today}

\maketitle

\section{Introduction}

Single-photon sources are widely developed worldwide \cite{Eisaman11}. Some produce single-photons in a heralded way, e.g. from a pair source, the detection of one photon heralding the production of its twin. Others produce them on-demand, e.g. using a single quantum emitter. But what for? \\

For fun? Certainly. Because physicists enjoys manipulating individual quanta. Fine. But, you may have noticed that this argument does not provide good enough motivations for research proposals. Applications, even hypothetical, are often necessary to obtain research funding. \\

To build up a quantum computer? Maybe. It has been shown that linear optical quantum computing could work if the product of the detector efficiency and the single photon emission probability is greater than 2/3 \cite{Varnava08} (and even greater than 1/2 if one uses photon pair sources \cite{Gong10}) provided that the sources never emit more than one photon, that the corresponding photon wave-packet is pure and that the emission is perfectly indistinguishable from one source to another. But maybe not. A more realistic study, taking the multi-pair emissions into account, has recently shown that very basic entangling gates require photon-detector and single-photon sources with efficiencies of about 0.9 and single-photon purities (g$^{(2)}$ second order auto-correlation function) better than 0.07 \cite{Jennewein11}. Note that these performance levels do not allow one to perform a useful calculation nor to show the superiority of quantum computation over its classical counterpart. They merely correspond to a guideline e.g. for the implementation of a heralded entangled photon pair.\\

For quantum communication tasks? Good question. Quantum cryptography has become the field of preference for many of us \cite{Gisin01}. This is an exciting research domain that  benefits from the more fascinating concepts lying at the intersection between quantum physics and information theory to securely exchange keys. However, as quantum cryptography becomes more and more mature, we now know what it requires and what it doesn't. For example, many developers of single-photon sources state that their device is useful for point to point quantum key distribution (QKD). But as we will see in the next section, high secret key rates can be achieved much more conveniently using weak laser pulses with appropriate protocols. \\

It is thus natural to wonder whether a source producing single photons is a useful resource. The aim of this paper is to highlight the central role that single-photon sources could play for future quantum communication. In section \ref{QKD}, we recall the requirements to make single-photon sources useful for standard quantum key distribution protocols and we conclude that faint laser pulses better do the job today and will likely provide superior performance for a long time. We show in section \ref{DI-QKD}, that single-photon sources may open the way for new kinds of QKD protocols, which exploit nonlocal correlations to make the security of the key independent of the device that is used to produce it. Interestingly, this would turn quantum nonlocality from a fundamental question to an applied physics concept. The price to pay is that the photons need to be created with high efficiencies, at high rates, in pure and indistinguishable states and at telecom wavelengths. We also recall that, in this context, the on-demand production of photons would greatly speedup the secret key rate. In  section \ref{QRep}, we focus on long distance quantum communication using quantum repeaters. We show how single-photon sources can be used to build up efficient architectures, significantly more efficient than protocols based on currently available photon-pair sources. We show that in addition to being in pure and indistinguishable states, the photons produced on-demand need to be compatible with the memory bandwidth to be useful in the framework of quantum repeaters. The last section is devoted to our conclusion. \\

\section{Requirements for QKD}
\label{QKD}

QKD enables two parties, Alice \& Bob, to share a random key known only by them, which can be subsequently used e.g. to communicate in a secure way \cite{Bennett84}. QKD protocols rely on the fascinating quantum property that information gain is impossible without introducing errors if the communication relies on non-orthogonal quantum states. In other words, the laws of Nature make QKD secure.\\

If you sketch QKD on a piece of paper, it is natural to consider a perfect single photon emitter as a light source. By doing this, the security analysis is rather simple as a potential eavesdropper, Eve, has no means to split the signal. However, if you are an experimentalist, you will quickly understand that the realization of a true single-photon source is not a piece of cake. You will be tempted to use strongly attenuated laser pulses $|\alpha\rangle$ as an approximation for single photons, choosing an average photon number per pulse $\mu=|\alpha|^2$ of less than 1, say 0.1. This approximation seems reasonable in the sense that the chance to find two or more photons in a pulse becomes small (about 0.5 $\%$ in this case). So, most early experiments of QKD were based on such faint laser sources without remorse with $\mu=0.1$. However, it was pointed out by Brassard et. al. \cite{Brassard00} that if the transmission of the quantum channel is actually smaller than the probability that a nonempty weak pulse contains more than one photon $|\langle \alpha | \sum_{n>1} n \rangle|^2/|\langle \alpha | \sum_{n>0} n \rangle|^2$, the scheme is no longer secure. Indeed, in this case an eavesdropper can perform the so called photon number splitting attack \cite{Brassard00, PNS}. This attack consists in removing one photon of the pulses containing more than one and sending the remaining through a lossless fiber to Bob. All other pulses are blocked in order to keep the expected count rate at Bob's location. The removed photons are stored in a quantum memory and measured once Alice \& Bob have agreed on the measurement basis during the sifting process. Therefore, Eve obtains full information about the key without creating errors. The straightforward way to rule out this attack consists in reducing the average number of photon per pulse $\mu$ accordingly to the transmission efficiency $t$ ($\mu = t$ has been shown to be optimum for the BB84 protocol \cite{Bennett84}). Needless to say, that this reduces drastically the achievable bit rates and limits significantly the maximum transmission distance. \\

However, one should not precipitously conclude that single-photon sources are mandatory for QKD. Indeed, there are a couple of clever countermeasures allowing one to implement efficient QKD with faint laser pulses. The most simple trick is a mere change in the sifting procedure of BB84. Instead of announcing the basis, Bob announces his meaurement result. Alice now knows when Bob did not measure in the right basis and the measurement basis can be used to form the raw key. Eve, on the other hand, does not know in which basis she should measure her photon, so she cannot retrieve full information. It has been shown that the corresponding optimum $\mu$ goes with $\sqrt{t}$ \cite{Scarani04}. A second option is the decoy state protocol \cite{Hwang03}, where Alice varies on purpose the intensity of the pulses. By acting differently on pulses with different photon numbers, Eve will alter the detection statistics of Bob and the photon-umber splitting attack can be revealed. Finally another strategy consists in checking the coherence between different qubits, which prevents Eve acting on individual qubits as it is the case in a photon-number splitting attack. The diffrential phase shift \cite{DPS} and the coherent one-way schemes \cite{COW} are based on this strategy. It turns out that these latter protocols allow one to use rather high $\mu,$ from 0.2 to 0.5 even for very long distances. \\

The question is now, can single-photon sources, so called photon guns (on demand) or heralded single-photon sources, at telecom wavelengths, compete with the cheap and still efficient faint laser sources? The answer is clearly no for the following reasons: 1) The collection efficiency of single photon sources is usually very small. However, even if one assumes a high collection efficiency of say 0.8, one has to take into account additional loss, e.g for rapid integrated phase or polarization modulations with typical loss of 2 dB or more. So finally, at the output of Alice's device, the average number of photon will not be higher than for faint laser sources. 2) Faint laser schemes run at repetition frequencies as high as 1 GHz. This means that single photon gun should feature jitter below 1 ns, but also almost Fourier limited bandwidth in order to limit dispersion and allow one for dense wavelength division multiplexing. This is actually not the case for any single-photon source at telecom wavelength. Heralded single-photon sources based on photon-pair sources cannot achieve such a rate as it would ask for photon counters with GHz count rates. 3) Admittedly, the two points mentioned above are technological and not fundamental limitations. If some technological progress is certainly possible, it will be at unafordable price. A DFB laser diode is available below 1000 dollars. This is out of reach for any single-photon source, even without cryogenic cooling. A much higher price without a significantly improved performance is of course unconceivable for a commercial product. \\

So, single photon sources are useless for today's QKD systems. However, looking further ahead, we will see below that they are definitively needed for more complex schemes like for device-independent QKD and for long-distance quantum communication based on quantum repeaters. \\

\section{Requirements for Device independent QKD}
\label{DI-QKD}

We have seen above that weak pulse QKD is perfectly secure. However, the proofs  are obviously only valid for correctly implemented systems. But how can one be sure that the implementation has no weakness?  Indeed, a weakness could have been introduced by the manufacturer by negligence or even on purpose. This is a major problem if one wants QKD to run obliviously between two black boxes installed in banks. The solution consists either to educate the bankers on quantum physics or to submit the black boxes to some certification agencies. However, it would be much more elegant if one could demonstrate security without any assumption on the actual devices. This has lead to the concept of device-independent QKD \cite{Barett05, Acin07}.\\

Device-independent QKD relies on entanglement based protocols \cite{Ekert91}. The principle of the latter is quite simple. When both Alice \& Bob perform measurements along the z direction (or along x) on the singlet state, they get locally random results but perfectly anti-correlated. By repeating the experiment several times and by choosing the measurement settings randomly among $\{\sigma_z, \sigma_x\},$ they each get a string of random bits. Keeping the results for which the measurement setting were identical only (and if one of them flips the bits), they get the same copy of a string of random bits, that is, a key. But how can one make the security device independent using entanglement based protocols? \\

Non-locality performs this trick. Indeed, a possible way to detect entanglement is to perform measurements whose correlations violate a Bell inequality. Importantly for QKD, the violation of a Bell inequality first ensures the presence of entanglement between Alice \& Bob but also forbids a third party to share quantum correlations with them, independently of any details about the Hilbert space dimension and the measurement devices. This opens an avenue for bankers, and for everyone who doesn't know anything about quantum physics, to communicate in a secure way. \\ 

An experimental implementation of device-independent QKD is still awaited since hitherto, all optical Bell tests are subject to the detection loophole. Indeed, all entangled photons are not detected because of unavoidable losses and the missed events could be used to perform powerful attacks. A potential eavesdropper could force the black boxes to produce results only if the measurement settings are in agreement with a predetermine scheme. Closing the so called detection loophole in an optical Bell test is therefore a requirement for the implementation of device-independent QKD. However, the detection efficiency (i.e. the probability that Bob gets an answer once he asks a question) that is required to rule out attacks based on the detection loophole is larger than 82.8\% for the CHSH inequality \cite{Clauser69} in the absence of other limitations. But the transmission efficiency of a 5-km-long fiber at telecom wavelength is roughly 80\%. Hence, transmission losses represent a fundamental limitation for the realization of a detection-loophole free Bell test on any distance relevant for QKD. The question that naturally follows is thus: how can we create entangled photon pairs at a distance in a heralded way? \\

\begin{figure}[ht!]
\includegraphics[width=6 cm]{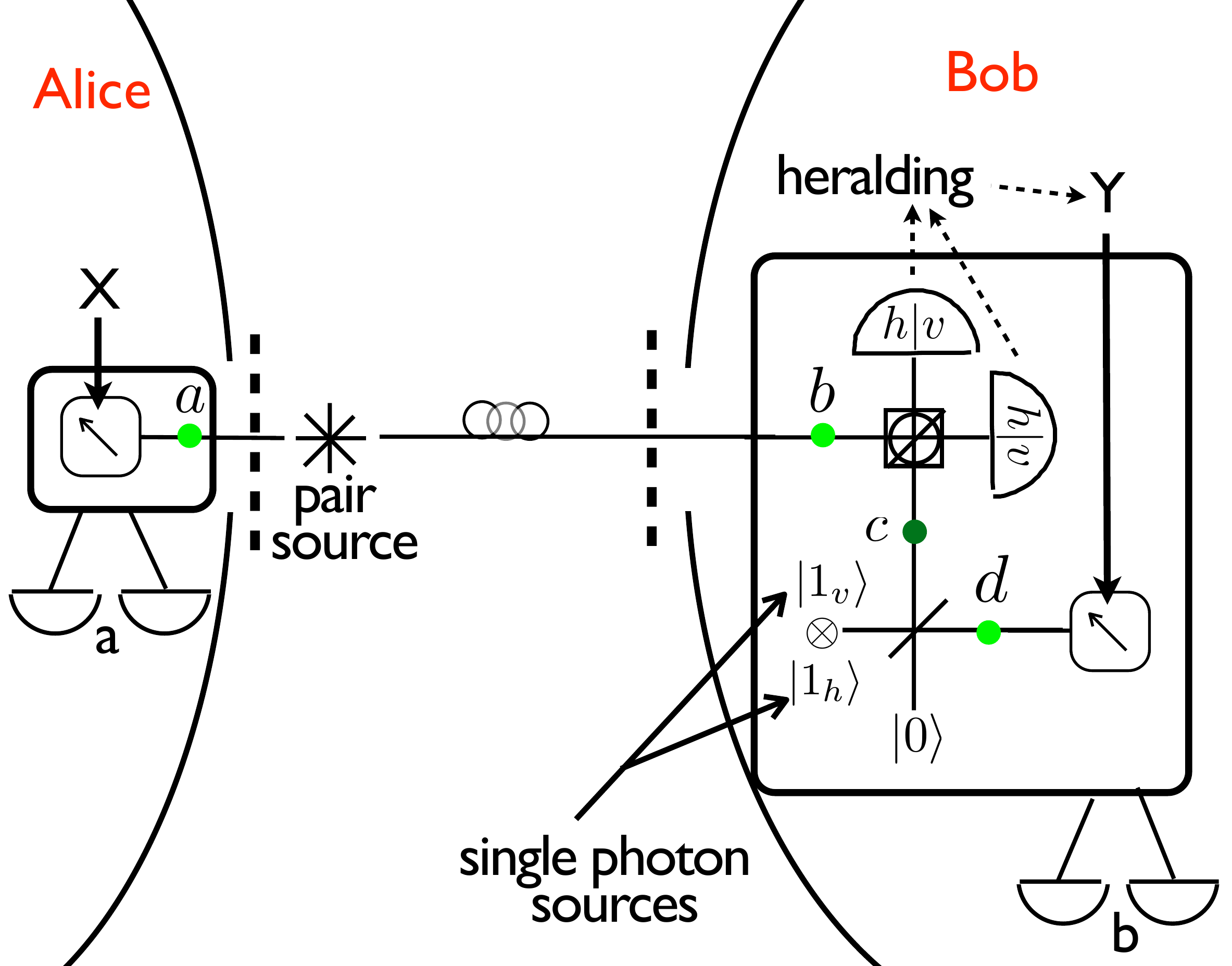}
\caption{Setup used to overcome the problem of transmission losses in a detection-loophole free Bell test. Star represents a photon source producing pairs, entangled in polarization, in mode $a$ and $b.$ Alice gets the mode $a.$ The mode $b$ is sent to Bob's location through an optical fiber. In order to know whether the photon successfully reaches its location, Bob's black box contains two single-photon sources. The photons that they produced have orthogonal polarizations, $h$ and $v$ respectively. They are sent into a beamsplitter to create entangled states involving the modes $c$ and $d$. The modes $b$ and $c$ are combined on a polarization beamsplitter in the $\pm45^\circ$ basis and subsequently measured in the $h|v$ basis (half-circles). The detected modes are  $D_\pm=c_h+c_v\pm b_h \mp b_v$ and $\tilde D_\pm = \pm c_h \mp c_v+b_h+b_v.$ The detection of one photon in mode $D_+$ and another one in $\tilde D_+$ project the mode $a$ and $d$ into a maximally entangled state provided that the mutli-pair emission probability is negligible and that the beamsplitter transmission probability is small enough. This holds even if the detectors have non-unit efficiencies and even if Alice \& Bob are far away from each other \cite{Gisin10, Minar11}. Therefore, this provides an operational protocol to herald the successful distribution of entanglement between far away locations. In the framework of device-independent QKD, Alice (Bob) chooses a measurement X (Y) which is applied on the mode $a$ $(d)$ and she (he) subsequently gets a result labelled a (b). Repeating the experiment several time, Alice \& Bob can access the probability distribution P(ab$|$XY) from which they can compute the CHSH value. The violation of the CHSH inequality insure Alice \& Bob that they share entanglement from which they can securely get a key. If Bob chooses a measurement only when he gets coincidence detections $D_+$ -- $\tilde D_+,$ the overall efficiency requires to rule out attacks based on the detection loophole reduces to the detection efficiency \cite{Gisin10}.}
\label{fig1}
\end{figure}

The solution that has been proposed in Ref. \cite{Gisin10} revolves around performing an entanglement swapping operation. Given two entangled photons, say the photon $a$ and $b,$ and another entangled pair $c$ and $d,$ it is possible to entangle $a$ and $d$ by performing a joint measurement of photons $b$ and $c$ in the Bell basis, provided that the result of the measurement is communicated to $a$ and $d$ \cite{Zukowski93}. The latter two end up entangled, even through they are located at remote location. Furthermore, the creation of their entanglement is heralded by the joint measurement. This is exactly what we were looking for. The only remaining question is how can one perform an entanglement swapping operation in practice? The most natural approach would be to use pair sources based on spontaneous parametric down conversion and linear optical elements to perform a partial Bell measurement. However, the emission of multi-pairs, inherent to spontaneous parametric down conversion, inevitably corrupt the Bell state measurement and drastically reduce the fidelity of the resulting entangled state.\\

\begin{figure}[ht!]
\includegraphics[width=8 cm]{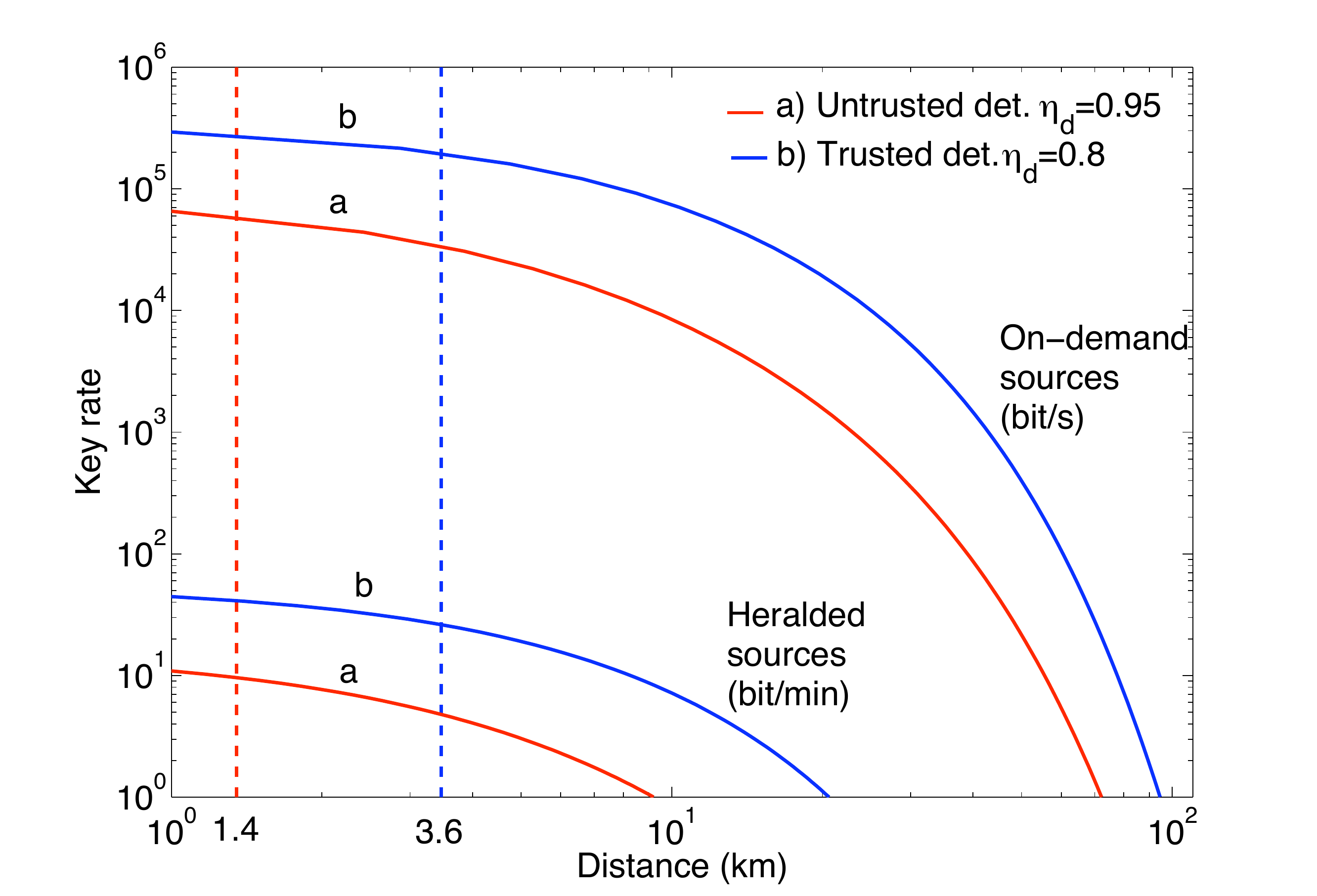}
\caption{Usefulness of on-demand single-photon sources for device-independent QKD. The quantity shown is the key rate for the given distance. (Red) curves labeled a correspond to a fully device independent implementation (detection efficiency 0.95). (Blue) curves labeled b correspond to the case where the detectors are trusted (detection efficiency 0.8), i.e. the eavesdropper cannot temper with them. The dotted vertical lines give the maximal distance that can be achieved without single-photon source (limited by the transmission efficiency). The two lower (upper) curves give the performance in bit per minute (bit per second) of the scheme shown in Fig. \ref{fig1} implemented with heralded (on-demand) single-photon sources. This figure comes from Ref. \cite{Gisin10}.}
\label{fig2}
\end{figure}

An attractive solution consists in using single-photon sources, because they never produce more than one photon. This is the central idea of the proposal presented in Ref. \cite{Gisin10} which consists in performing a conventional entanglement swapping operation but with one entangled-photon-pair source replaced by two heralded single-photon sources producing one photon each that is subsequently sent to a beamsplitter. As a consequence, the state resulting from the joint measurement is maximally entangled, independent of the loss, provided that the reflection of the beamsplitter and the probability that the pair source produces multi-photon pairs are low enough, c.f. fig. \ref{fig1} for details. If Bob chooses a measurement only when he gets a successful joint measurement, the overall efficiency requires to rule out attacks based on the detection loophole reduces to the detection efficiency. \\

By considering an implementation with sources based on parametric down conversion only, producing pure and indistinguishable photons, with a repetition rate of 10 GHz, a fiber attenuation of 0.2 dB/km corresponding to telecom wavelength photons, coupling efficiencies into fiber of 0.9, detection efficiencies of 0.8, this proposal achieves a key rate of a few bits per minute on a typical distance of 10 km. Note that if the photons emitted from the single-photon sources are produced on-demand, the achievable key rate is increased by several orders of magnitude. The usefulness of on-demand single-photon sources in the framework of device-independent QKD is well illustrated by the figure \ref{fig2}. \\

We will now see that single-photon sources could also play a central role for quantum repeaters.\\

\section{Requirements for quantum repeaters}
\label{QRep}

\begin{figure}[ht!]
\includegraphics[width=8 cm]{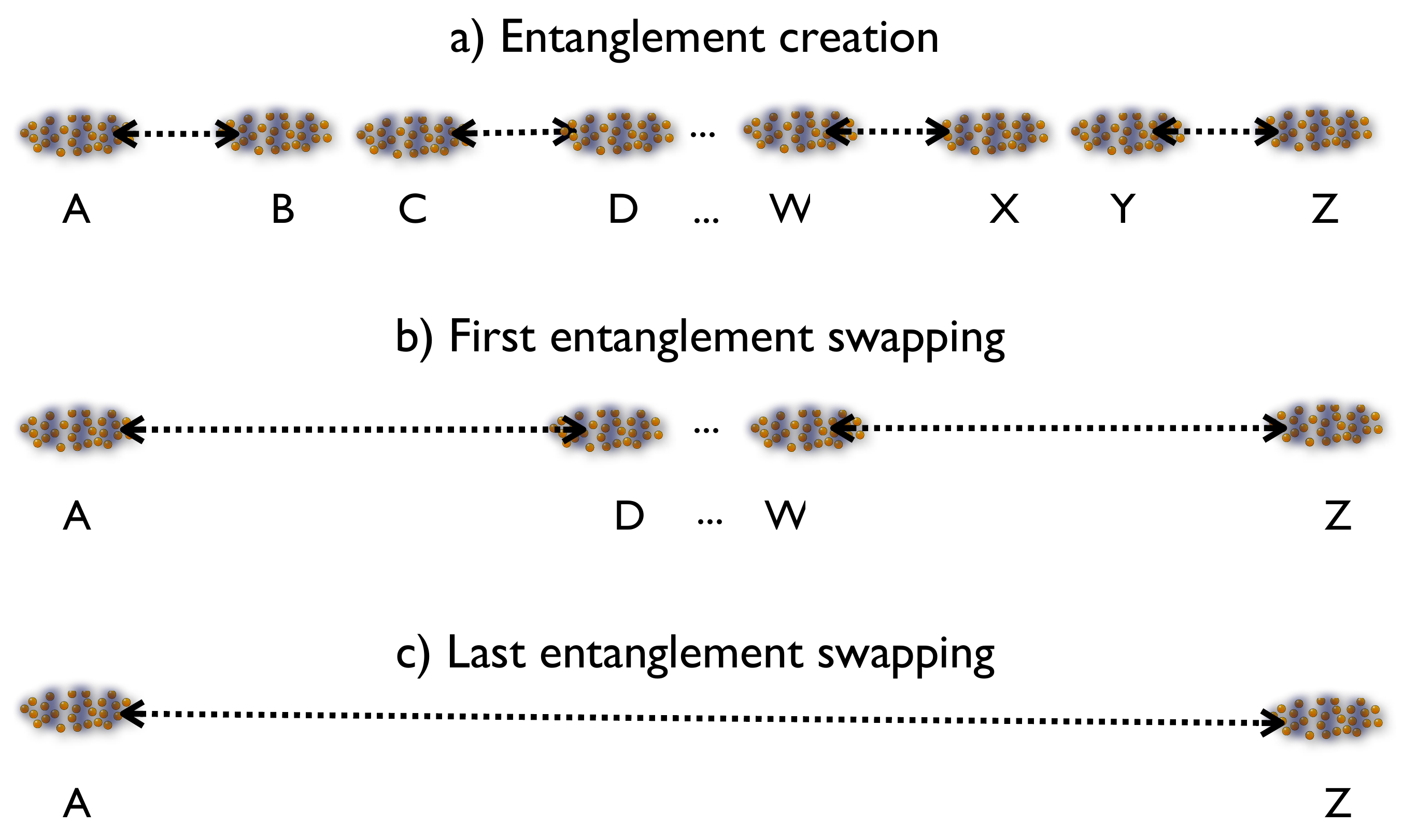}
\caption{Principle of quantum repeaters. The entanglement is first created and stored in each elementary link. Entanglement swapping operations are then repeatedly performed between neighboring links until it is extended over the desired distance.}
\label{fig3}
\end{figure}

The distribution of quantum states over long distances is limited by photon loss. For example, the transmission efficiency of a 1000 km long optical fiber is $10^{-20}$ \cite{footnote_loss}. Even under a very optimistic assumption on the repetition rate of the source, say 10 GHz, this translates into the distribution of one quantum state every 300 years on average! Moreover, the problem of loss cannot be overcome by straightforward  amplificationn as in classical communication, because of the no-cloning theorem \cite{Wootters82}. But let us keep hope alive!\\

Briegel et al. \cite{Briegel98} proposed an attractive solution that relies on the amazing properties of entanglement, the so-called quantum repeater. The principle is sketched in Fig. \ref{fig3}. It consists in dividing the overall distance into elementary links, distributing the entanglement in each of the links, and then swapping repeatedly the entanglement between neighboring links until it is extended over the desired distance. If $\eta_t$ is the transmission efficiency within one link and if the quantum repeater of interest is made with N links, then the average rate with which entanglement is distributed scales as $\eta_t$ for quantum repeaters whereas it scales as $\eta_t^N$ for the direct transmission of photons through an optical fiber. However, the price to pay is that first, the distribution of entanglement in each link needs to be heralded. Furthermore, the entanglement has to be stored in each link and processed many times to be swapped between adjacent links. \\

\begin{figure}[ht!]
\includegraphics[width=8 cm]{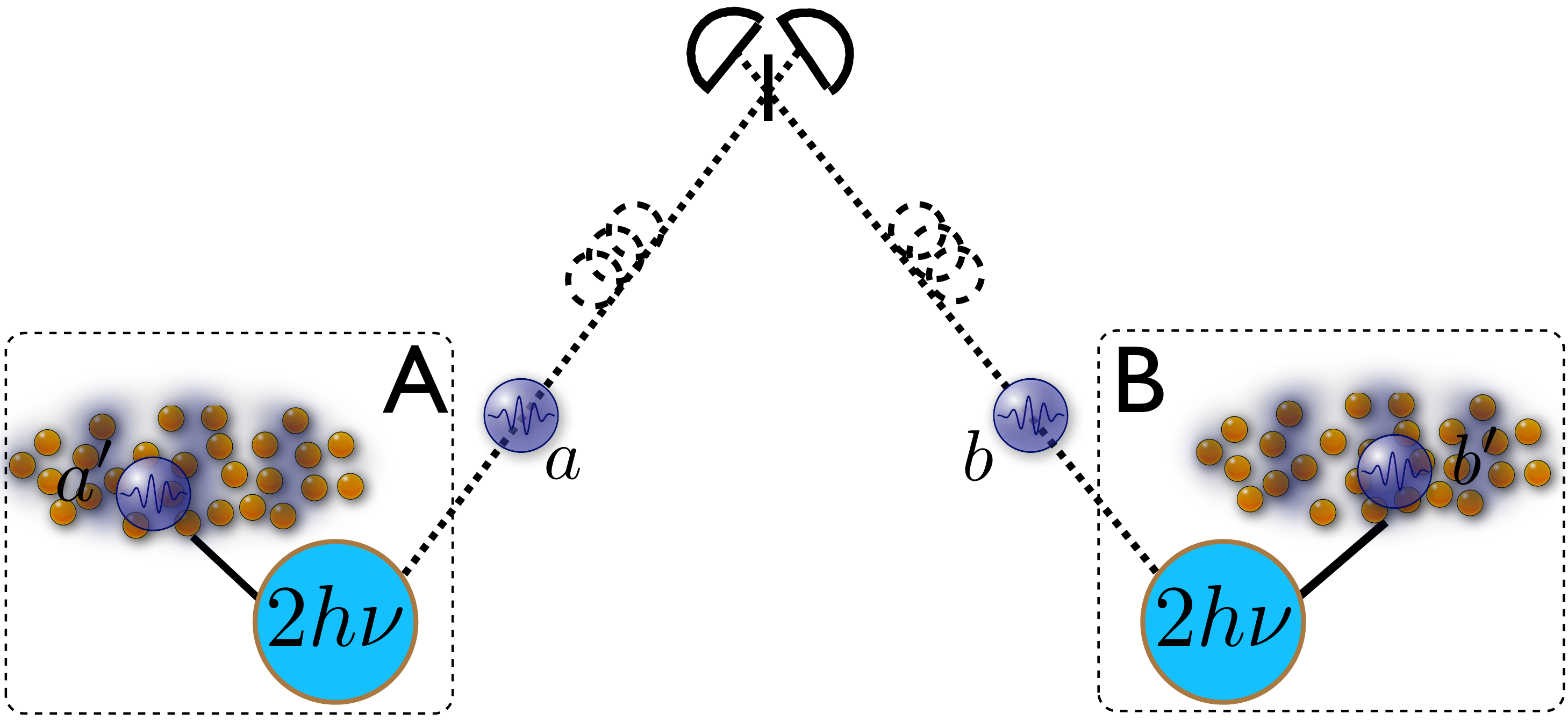}
\caption{Heralded entanglement creation within an elementary link which is inspired by the DLCZ protocol \cite{Duan01}. Photon-pair sources, producing two-mode squeezed states (as obtained with spontaneous parametric down conversion in non-linear crystal) are coherently excited so that one of them emit a photon pair. The mode $a'$ and $b'$ are stored locally in atomic ensembles, whereas the modes $a$ and $b$ are sent to a central station where they are combined into a beamsplitter. The detection of one photon after the beamsplitter heralds the storage of a single excitation delocalized among the two crystals, i.e. the entanglement between the two memories. The entanglement can then be swapped by releasing the stored photons using linear optics and photon counting techniques \cite{Simon07}.}
\label{fig4}
\end{figure}

Duan and co-authors \cite{Duan01} (DLCZ) have shown how to meet all the above requirements using linear optical elements for the entanglement processing operations and atomic ensembles for the quantum memories \cite{Sangouard11}. It basically requires one photon pair source and one memory at each location. The two sources are coherently excited so that the detection of a single photon which could have been emitted from either of two sources, heralds the entanglement between the two memories, c.f. Fig \ref{fig4}. However, because of the rapid growth (quadratic with the number of links) of errors due to multiple emissions during the entanglement swapping operations, it is necessary to work with low emission probabilities, i.e. to weakly excite the pair sources so that the multi-pair emission is negligible \cite{Sangouard08}. This inevitably decreases the achievable entanglement distribution rate.\\

\begin{figure}[ht!]
\includegraphics[width=8 cm]{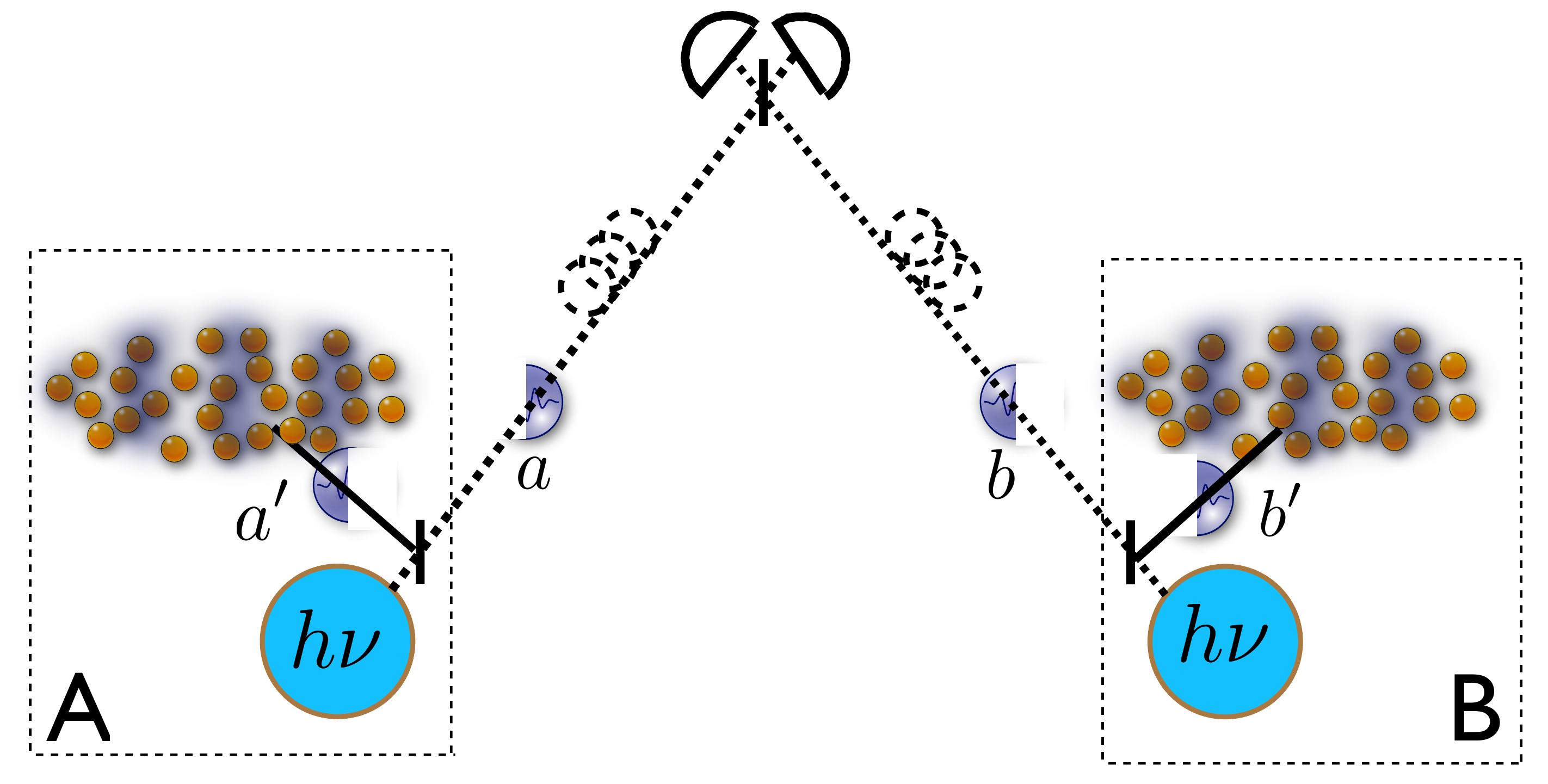}
\caption{Heralded entanglement creation within an elementary link for the protocol of Ref. \cite{Sangouard07} which is based on single-photon sources. The two on-demand sources produce one photon each, subsequently sent into a beamsplitter. This creates entanglement between the modes $a'$-$a$ and $b'$-$b.$ The modes $a'$ and $b'$ are stored locally. The detection of one photon at the central station where the modes $a$ and $b$ are combined into a beamsplitter, heralds the creation of entanglement between the remote memories (i.e. a superposition of memory A and B being excited). The swapping operations are identical to the ones used in the DLCZ protocol.}
\label{fig5}
\end{figure}

The use of single-photon sources, to eliminate errors due to two pair emissions, leads to a significant improvement in the achievable entanglement distribution rate. Let us focus on an example whereby the distribution of entangled pairs over 1000 km and consider detector and memory efficiencies of 0.9. If one demands that the overlap between the distributed state and the singlet state is at least equal to 0.9, it can be shown that a protocol based on single-photon sources (c.f. Fig. \ref{fig5}) with source efficiencies of 0.95 and two photon emission of the order of $10^{-4},$ achieves entanglement distribution rates 18 times greater than the DLCZ protocol \cite{Sangouard07} . Note that efficient sources are required to take advantage of the protocol of Ref. \cite{Sangouard07}. More precisely, the single-photon source protocol of Ref. \cite{Sangouard07} achieves an advantage over the DLCZ scheme as soon as the single-photon source efficiencies are larger than 0.67. \\

We emphasize that single-photon sources can be exploited to herald the production of high fidelity entangled photon pairs from a probabilistic pair source based e.g. on spontaneous parametric down conversion, as shown in Fig. \ref{fig1}. It has also been shown in Ref. \cite{Minar11} how the potential imperfections affecting the states created from this source caused e.g. by detectors with non-unit efficiencies, can be systematically purified from an entanglement swapping operation, the later being used for entanglement creation within the elementary links. Near perfect fidelity entangled states are thus distributed in each link and this fidelity is preserved when the entanglement is extended over long distances through appropriate entanglement swapping operations. The resulting entanglement distribution rate is three orders of magnitude greater than the one corresponding to the DLCZ protocol \cite{Duan01}. The architecture presented in Ref. \cite{Minar11}  fully exploits the properties of single-photon sources and is currently the most efficient quantum repeater known to us based on atomic ensembles and linear optics.\\

The assumptions behind quantum repeater protocols based on single-photon sources, is that the photons are produced in pure and indistinguishable states, with a very high efficiency (typically higher than 70\%) and with a negligible pair contribution ($\approx 10^{-4}$ for \cite{Sangouard07}). Futhermore, they need to be narrowband in order to be compatible with the memory bandwidth (at most of the order of hundreds of MHz). The wavelength is also important but frequency conversion techniques \cite{conversion} can be used to bridge the gap between quantum memory and telecommunication wavelengths. \\

\section{Conclusion}

Single-photon sources never produce more than one photon at a time. However, present day quantum cryptographers could not care less because faint lasers are efficient, cheap and easy to use. Nevertheless, 
QKD would gain more users' trust by making the security independent of the devices. Single-photon sources  could play a key role in the first implementations of device-independent quantum key distribution which pleases everybody, both non-physicists and post-quantum cryptographers. In parallel, there is a need to extend the range of quantum communication. Here also, single-photon sources could well be the main ingredient for the realization of the first long-distance quantum communication based on quantum repeaters. Time will tell. \\ 

We thank Ci Wen Lim, Clara Osorio, Bruno Sanguinetti and Rob Thew for interesting discussions. We gratefully acknowledge support by the EU projects Qessence and Qscale, and from the Swiss NCCRs QP and QSIT.


\begin{thebibliography}{3}
\bibitem{footnote_loss} The loss of a typical telecom optical fiber is of 0.2 dB/km for the optimal wavelength around 1.5 $\mu m$.
\end{thebibliography}
\end{document}